# Gibbs free energies of Fe clusters can be approximated by Tolman correction to accurately model cluster nucleation and growth

Alexander Khrabry[1], Louis E.S. Hoffenberg[1], Igor D. Kaganovich[2], Yuri Barsukov[2], David B. Graves[1]

[1]*Princeton University, Princeton, NJ, USA*

[2]*Princeton Plasma Physics Laboratory, Princeton, NJ, USA*

**Abstract**

Accurate Gibbs free energies of Fe clusters are required for predictive modeling of Fe cluster growth during condensation of a cooling vapor. We present a straightforward method of calculating free energies of cluster formation using the data provided by molecular dynamics (MD) simulations. We apply this method to calculate free energies of Fe clusters having from 2 to 100 atoms. The free energies are verified by comparing to an MD-simulated equilibrium cluster size distribution in a sub-saturated vapor. We show that these free energies differ significantly from those obtained with a commonly used spherical cluster approximation – which relies on a surface tension coefficient of a flat surface. The spherical cluster approximation can be improved by using a cluster size-dependent Tolman correction for the surface tension. The values for the Tolman length and effective surface tension were derived, which differ from the commonly used experimentally measured surface tension based on the potential energy. This improved approximation does not account for geometric magic number effects responsible for spikes and troughs in densities of neighbor cluster sizes. Nonetheless, it allows to model cluster formation from a cooling vapor and accurately reproduce the condensation timeline, overall shape of the cluster size distribution, average cluster size, and the distribution width. Using a constant surface tension coefficient resulted in distorted condensation dynamics and inaccurate cluster size distributions. The analytical expression for cluster nucleation rate from classical nucleation theory (CNT) was updated to account for the size-dependence of cluster surface tension.

1. Introduction

Iron nanoparticles have multiple applications in medicine, diagnostics, and environmental projects (Govardhane. and Shende 2022; Kapoor et al. 2024; Pugazhendhi et al. 2018; Xu et al. 2022). A prominent application is catalysis in floating-catalyst chemical vapor deposition (FCCVD) synthesis of carbon nanotubes (Hoecker et al. 2016; Hussain et al.2018; Kim et al. 2014a). In FCCVD reactors, catalytic nanoparticles can be produced directly in the reactor from condensation of Fe vapor (Hoecker et al. 2016, 2017; Hussain et al.2018; Kim et al. 2014a, Sehrawat et al. 2024). Control over the size distribution of the forming catalyst nanoparticles is crucial for the synthesis of carbon nanotubes with desired characteristics, such as purity, number of walls, diameter, length, crystallinity, and selective chirality (Nessim et al. 2008; Yu et al. 2012). Accurate predictive models of Fe condensation are required to control catalyst nanoparticle formation and growth (Bilodeau and Proulx 1996), and therefore the carbon nanotube product.

Multiple kinetic solvers and simplified moment models are available to model formation and growth of liquid particles (clusters) from a condensing vapor (Bilodeau and Proulx 1996; Colombo et al. 2012; Frenklach and Harris 1987; Frenklach 2002; Friedlander 1983; Khrabry et al. 2024; Prakash, Bapat, and Zachariah 2003; Shigeta and Murphy 2011; Tacu, Khrabry, and Kaganovich 2020). All of these models require particle thermodynamic data to determine rates of reverse kinetic processes. The forward processes are particle growth through atom attachment of atoms, and reverse rates are atom evaporation from the particle surface. Most commonly, the spherical cluster approximation is used (Girshick and Chiu 1990; Girshick, Chiu, and McMurry 1990) to determine Gibbs free energies of all particles – where the nascent clusters consist of several atoms, and the grown nanoparticles comprise millions to billions of atoms. This approximation forms the basis of the classical nucleation theory (CNT; Bakhtar et al. 2005; Girshick 2024; Katz 1977; Smirnov 2000,2010) which predicts the rate of nucleation of thermodynamically stable clusters, (i.e., clusters that exceed the so-called critical size) (Frenkel 1955, Smirnov 2006).

In the spherical approximation, Eq. (7), the Gibbs free energy of a cluster formation is derived from the Gibbs energy of bulk liquid (consisting of the same number of atoms) by adding a surface energy correction term. This correction term is simply defined as the cluster's surface area times the surface tension coefficient. Typically, a constant surface tension coefficient for a flat liquid surface is used. However, the accuracy of this approximation for Gibbs free energies of small clusters is dubious, and its effect on the modeling of cluster formation from condensing vapor has not been thoroughly studied.

Time-of-flight (TOF) mass spectra experiments (Sakurai et al. 1998, 1999) showed that the stability of small $Fe_i$ clusters, consisting of several atoms to tens of atoms, is a non-monotonic function of the cluster size. Clusters $i$ = 7, 13, 15, 19 and 23 are commonly referred to as 'magic numbers' due to their higher stability and abundance. This observation implies that the free energies of Fe cluster formation are substantially different from those predicted by the spherical approximation, at least for $i ≤ 100$ atoms. *Ab-initio* density functional theory (DFT) modeling of Fe clusters (Kim et al. 2014b) revealed 'magic numbers' in the binding energies of solid Fe clusters in the ground state (at 0 K). 'Magic numbers' were also observed in the magnetic (Aktürk and Sebetci 2016; Ma et al. 2007) and melting (Hoffenberg et al. 2024) properties of small Fe clusters.

There are ongoing debates about whether the effects of cluster 'magic numbers' and non-sphericity should be accounted for in condensation models. The argument against including 'magic numbers' is that liquid clusters have a disordered structure, which may smoothen out the geometric effects of atom packing (Smirnov 2010).

Girshick, Agarwal, and Truhlar (2009) have analyzed Gibbs free energies of formation of molten Al clusters up to 60 atoms in size. These Gibbs free energies were obtained (Li et al. 2007) using quantum chemistry modeling and kinetic Monte-Carlo methods to scan the cluster configuration space. They have concluded that the Gibbs energies differ both qualitatively and quantitatively from the spherical approximation. Spikes and troughs were observed in the Gibbs energy values as a function of cluster size, corresponding to 'magic numbers' of atoms in a cluster. This should result in a qualitatively different nucleation rate than that predicted by CNT, and should affect the overall condensation process. Unfortunately, to the best of our knowledge, Gibbs free energies of formation of molten iron clusters needed for the kinetic modeling of cluster growth have not been calculated.

Note that, for the metals in the first, second, and third columns of the periodic table, the 'magic numbers' are associated with cluster electronic structure. Quantum chemistry approaches can accurately model this effect. In addition, the jellium model (De Heer et al. 1987) – reminiscent of the Thomson (plum pudding) model of atoms and based on the assumption that the metal clusters consist of the electron gas weakly interacting with a positive background ('jellium') – predicts the appearance of the 'magic numbers'. The metal clusters can be modeled as a superatom (Bergeron et al. 2004, Castleman and Khanna 2009), where electrons occupy the $1s^2 1p^6 1d^{10} 2s^2 1f^{14} 2p^6$... energy levels. The clusters with filled electronic shells are less reactive and consequently more stable. This 'electronic magic number' effect should be the strongest for alkali metal clusters (Knight et al. 1984; Göhlich et al. 1991), for which the 'magic numbers' were observed experimentally in clusters up to thousands of atoms (Knight et al. 1984, Martin et al. 1994, Martin 1996). For transition metals, like Fe, the effect of electronic structure on Gibbs energies is less pronounced. Sakurai et al. (1999) found that the magic numbers of $Fe_i$ clusters (i = 7, 13, 15, 19) correspond to the pentagonal bipyramid, icosahedron, bcc structural unit, poly icosahedra, respectively. The major contribution to the Gibbs energies originates from the geometric effects of atomic configuration in a cluster. Thereby, classical potentials might be used to approximate Gibbs free energies of Fe clusters.

The free energies of formation of $CO_2$ clusters at three temperatures have been retrieved from classical molecular dynamic (MD) simulations in Halonen et al. (2021), by direct observation of the change in cluster number density during formation and evaporation. It was shown that CNT can be improved by introducing a Tolman-like term into the spherical model expression for the free energy of formation of $CO_2$ clusters.

This work proposes an alternative, straightforward, and economical method to determine Gibbs free energies of liquid clusters as a function of temperature using data directly accessible from MD simulations of individual clusters. We apply this method to determine Gibbs free energies of formation of Fe clusters from 2 to 100 atoms using classical MD simulation data from the accompanying paper (Hoffenberg et al. 2024). The resulting free energies are compared to those from the spherical cluster approximation. These free energies are then applied in an efficient kinetic solver (Khrabry et al. 2024) to model the evolution of the entire cluster size distribution during condensation of a rapidly cooling Fe vapor.

The paper is organized as follows. Section 2 describes the method of calculating cluster free energies from MD and how the kinetic solver uses these energies to model cluster size distribution. The results of this research are presented and discussed in Section 3. The free energies of Fe cluster formation are presented in Section 3.1 and benchmarked in Appendix 4. The difference between the accurate free energies of the Fe clusters and the spherical cluster approximation is analyzed in Section 3.2. In Section 3.3, the effect of cluster free energies and their approximations on the modeling of cluster formation is analyzed. In Section 3.4, we present an expanded analytical expression for cluster nucleation rate and analyze its accuracy by comparing its predictions to the numerical solutions. And, finally, Section 4 contains the summary of the work.

## 2. Methods

**2.1. Determining cluster Gibbs free energies from classical MD**

This section describes a method for obtaining Gibbs free energies of cluster formation from easily accessible MD modeling software: 1) average kinetic and potential energies of individually modeled clusters, and 2) the Hessian (or second derivative) matrix for the cluster energies with respect to atomic displacements.

This work utilizes caloric curves of Fe clusters of 2 to 100 atoms obtained and presented in detail in our previous work (Hoffenberg et al. 2024). The caloric curves represent the dependence of the clusters' total energy on temperature $T$. Each datapoint in each of the caloric curves was obtained in a separate 10 ns microcanonical ensemble (NVE) MD simulation of an individual cluster. The simulations were performed in LAMMPS (Thompson et al.) using the embedded atom method Finnis-Sinclair potential (Finnis and Sinclair 1984; EAM-FS). This potential was parameterized to reproduce solid and liquid characteristics of Fe (Mendelev et al. 2003). The results of the classical MD are benchmarked by comparison with ab-initio (DFT-based) MD modeling of the $Fe_{13}$ cluster and binding energies of various Fe clusters at the ground state in Appendix 2. As an example, a caloric curve for $Fe_{13}$ is shown in Appendix 2.

The MD simulations were initiated with no whole-cluster rotational or translation motion (all cluster's energy was distributed within its internal degrees of freedom $i_{int}$). Thereby, temperature $T$ of a cluster could be determined from its time-average kinetic energy $E_k$ as follows:

$$T = \frac{2E_k}{i_{int}k}, \qquad (1)$$

where $i_{int} = 1$ for a dimer, and $i_{int} = 3i - 6$ for an $i$-atom cluster with $i>2$; $k$ is the Boltzmann constant.

Cluster's total energy is simply the sum of its potential and kinetic energies:

$$E = E_k + E_p. \qquad (2)$$

The Gibbs free energy of a cluster formation can be expressed through its enthalpy and entropy of formation:

$$\Delta G_0^f = \Delta H^f + T\Delta S_0^f. \qquad (3)$$

The superscript *f* denotes cluster formation from gas phase atoms; subscript 0 denotes atmospheric pressure conditions. Note: the enthalpy of formation $\Delta H^f$ of clusters of a particular size is only a function of temperature (not pressure), hence no subscript 0. Gibbs free energy refers to the energy of a mol of gas of a given sort of clusters (not a single cluster, as $E_k$ and $E_p$).

Enthalpy of cluster formation can be immediately obtained from its total energy by adding energies of rotational and translational degrees of freedom and the p-V work:

$$\Delta H^f = N_A E + \left(\frac{i_{add}}{2} + 1\right)RT. \qquad (4)$$

Here, $i_{add}$ is the number of additional degrees of freedom that need to be accounted for a gas consisting of one sort of clusters. $i_{add} = 5$ for a dimer, and $i_{add} = 6$ for bigger clusters. Three degrees of freedom (DoF) come from cluster rotation (two DoF for a dimer), three DoF account for cluster's translational motion, and the extra $RT$ term accounts for the PV work. $R$ is the gas constant ($R = kN_A$, where $N_A$ is Avogadro's number). Note that the variables $E$, $E_k$ and $E_p$ describe individual clusters and are measured in *J*, whereas the thermodynamic variables $\Delta G_0^f$, $\Delta H^f$ and $T\Delta S_0^f$ describe cluster gas and are measured in J/mol.

The formation entropies $\Delta S_0^f$ are more complicated to compute. The same caloric curves data can be utilized to integrate over temperature to identify enthalpies from their thermodynamic definition:

$$\Delta S_0^f(T) = \Delta S_0^f(T_0) + \int_{T_0}^{T} \frac{d(\Delta H^f)}{T}. \qquad (5)$$

This expression shows that even though each datapoint of the caloric curve is obtained in a separate simulation, the whole curve is representative of the cluster's thermodynamic properties. It also describes cluster melting, as addressed in detail in Hoffenberg et al. (2024). Datapoints in the caloric curves by Hoffenberg et al. (2024) have a step of about 50 K which is sufficiently frequent to accurately perform this numerical integration.

The initial entropy value $\Delta S_{0,T_0}^f$ can be found using an initial temperature $T_0$ that is low enough for a harmonic approximation for solid cluster oscillations to be used. A detailed process of determining the entropy of solid clusters in a gas phase is described in the documentation of the VASP DFT code (Kresse and Hafner 1993) and Appendix 1.

Once the Gibbs energies of cluster formation $\Delta G_0^f$ are obtained for the atmospheric pressure, they can be recalculated for an arbitrary pressure. Since $\Delta H^f$ does not depend on pressure, and the dependence of $\Delta S_0^f$ on pressure is given by Eq. (A1.11), the expression for the free energies at $\Delta G_f$ at an arbitrary pressure $p$ reads:

$$\Delta G^f = \Delta G_0^f - (i-1)RT \ln \frac{p}{p_0}. \tag{6}$$

**2.2. Determining effective surface tension coefficient**

As was mentioned in the Introduction, in the spherical approximation commonly used in CNT (Girshick and Chiu), Gibbs free energy of formation of an *i*-atom cluster is defined as:

$$\Delta G_i^{f,spher} = -(i-1)RT \ln S + \left(i^{\frac{2}{3}} - 1\right) 4\pi N_A r_W^2 \sigma. \tag{7}$$

Here, the first term in the right-hand side (RHS) represents the change of the Gibbs free energy upon condensation of $i$ atoms into the bulk liquid; the second term in the RHS represents the effect of the cluster's surface.

Note that this approximate 'ad-hoc formula' was not derived from a strict theory. For example, it does not account for the rotational degrees of freedom of a cluster. Also, both terms in the RHS depend on the number of atoms minus one (i.e., $i - 1$ and $i^{\frac{2}{3}} - 1$ instead of simply $i$ or $i^{\frac{2}{3}}$). The original formula from CNT (Katz 1977), used the number of atoms $i$ instead, but was not self-consistent (Girshick and Chiu 1990). It predicts a non-zero energy of formation for a monomer from itself. At this point, Eq. (7) is the most commonly accepted form of the spherical cluster approximation for the cluster free energy.

In Eq. (7), $r_W$ is the Wigner-Zeitz radius of the liquid, $\sigma$ is the surface tension coefficient, $S$ is the vapor saturation degree defined as the ratio of pressure of the vapor $p$ (or density $n_1$) to the equilibrium vapor pressure $p_e$ (or density $n_1^e$) at a given temperature $T$:

$$S = \frac{p}{p_e} = \frac{n_1}{n_1^e}. \tag{8}$$

The equilibrium vapor density $n_1^e$ is a function of temperature $T$. It is defined as density of vapor in equilibrium with a flat liquid surface and is given by the Clapeyron-Clausius equation:

$$n_1^e = \frac{p_0}{kT} \times exp\left(\frac{L}{R}\left(\frac{1}{T_{boil}} - \frac{1}{T}\right)\right), \tag{9}$$

where $L$ is the latent heat of evaporation of iron (in J/mol), and $T_{boil}$ is its boiling temperature at atmospheric pressure $p_0$.

If the Gibbs free energies of cluster formation at a given temperature and atmospheric pressure $\Delta G_0^f$ are known, then the effective surface tension coefficient $\sigma_{eff,i}$ can be derived from Eqs. (6)-(8):

$$\sigma_{eff,i} = \frac{(i-1)^{-\frac{2}{3}}\Delta G_0^f + (i-1)^{\frac{1}{3}}L\left(\frac{T}{T_{boil}} - 1\right)}{4\pi N_A r_W^2}. \tag{10}$$

For the MD iron system, the latent heat $L$ and boiling temperature $T_{boil}$ in Eqs. (9) and (10) are determined by the EAM-FS potential of Fe atoms and may differ from experimental values. Thereby, in order to make the definition of the effective surface tension self-consistent, we have performed a series of direct MD liquid-vapor coexistence simulations to regress the parameters $L$ and $T_{boil}$ corresponding to the EAM-FS potential.

In this series of direct vapor-liquid coexistence MD simulations, temperature was varied from 3000 K to 4500 K. For each temperature, the simulation was initiated by thermally equilibrating 2000 atoms in a liquid cube in a 10 ns-long simulation with periodic boundaries, as is shown in Fig. 1. The simulation box exceeded the size of the cubic slab by a factor of 5 in one direction to include five times the liquid volume of vacuum. A 1 fs timestep was used; constant pre-set temperature was maintained by a CSVR (canonical stochastic velocity rescaling; Bussi, Donadio, and Parrinello 2007) thermostat with a relaxation time of 100 fs. The final configurations after the initial thermalization step were used in the direct coexistence simulations, which were 50 ns NVE (microcanonical ensemble) simulations at the equilibrated temperature. Throughout the coexistence simulations, particles from the liquid slab desorb into the gas phase (former vacuum region) and reabsorb into the slab to establish equilibrium between the liquid and vapor phases. A time-averaged particle density ($\rho_{vap} = \frac{N_{vap}}{V}$) was calculated in the vapor region in each coexistence simulation, which was then used to calculate vapor pressure using the ideal gas law $P_{vap} = \frac{Nk_BT}{V} = \rho_{vap}k_BT$.

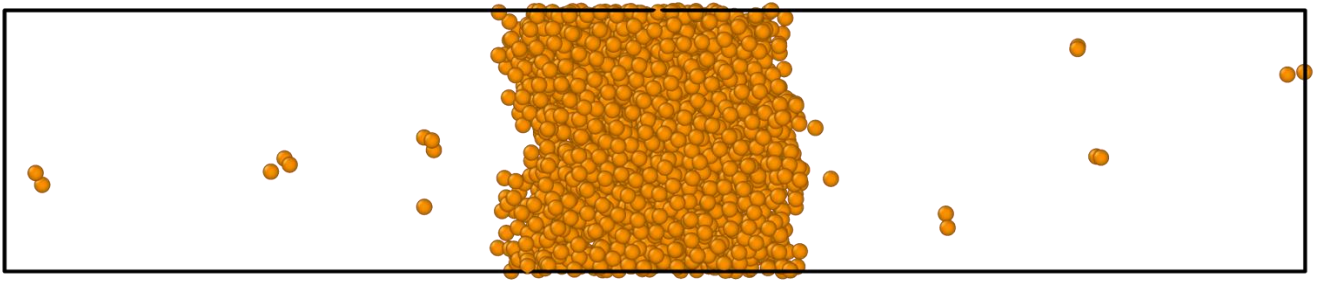

Fig. 1. Snapshot of a vapor-liquid coexistence simulation at 4000K. Atoms evaporate from the liquid slab in the center of the box to establish an equilibrium between the vapor and liquid phases.

**2.3. Modeling the cluster size distribution in a condensation process**

To model the evolution of the size distribution of particles forming in a condensing iron vapor, we use the particle kinetics solver from Khrabry et al. (2024). In this solver, the cluster growth kinetics are explicitly resolved for clusters of up to 100 atoms, as described by the kinetic equation (11), a.k.a. the general dynamics equation (GDE) or the particle balance equation (PBE). For bigger clusters, the size distribution is numerically approximated using an exponentially expanding computational mesh with an expansion factor of 1.2 and a high-accuracy numerical scheme.

The cluster growth kinetics are described by the kinetic equation:

$$\frac{dn_i}{dt} = f_{i-1}n_{i-1} - f_i n_i - r_i n_i + r_{i+1}n_{i+1} + \frac{dn_i}{dt}\bigg|_{coag} \qquad (11)$$

Here, $n_i$ stands for number density of clusters of *i* atoms (the equation is solved directly for all consecutive cluster sizes from 2 to 100 atoms). The first two terms in the right-hand side (RHS) represent rates of formation and destruction of *i*-atom clusters through atom attachment to (*i*-1)-atom clusters and atom evaporation from *i*-atom clusters. Similarly, the third and the fourth term in the RHS represent cluster attachment to *i*-atom clusters and atom evaporation from (*i*+1)-atom clusters. The last term in the RHS represents in a symbolic form effects of cluster collisions leading to cluster coagulation. $f_i$ and $r_i$ are forward and reverse rate constants defined as follows:

$$f_i = \frac{v_{th}}{4} n_1 s_{i-1}, \quad r_i = \frac{v_{th}}{4} s_{i-1} n_1^e \frac{n_{i-1}^e}{n_i^e}. \tag{12}$$

Here, $n_1$ is density of vapor atoms, $v_{th}$ is thermal velocity of vapor atoms, $s_i$ is the surface area of a cluster (surface shape is assumed), $n_1^e$ and $n_i^e$ are equilibrium densities of monomers and clusters of size *i*, respectively, at a given temperature.

The ratio of equilibrium cluster densities in Eq. (12) can be determined from the detailed balance equation:

$$\frac{n_{i-1}^e}{n_i^e} = exp\left(\frac{\Delta G_i^f - \Delta G_{i-1}^f}{kT}\right), \tag{13}$$

where $\Delta G_{0,i}^f$ and $\Delta G_{0,i-1}^f$ are Gibbs free energy of formation of an *i*-atom and (*i*–1)-atom clusters, as defined in Eq. (3).

This set of equations demonstrates why the Gibbs free energies of cluster formation are important for modeling the evolution of the cluster size distribution.

## 3. Results and discussion

### 3.1. Gibbs free energies of cluster formation

The effect of temperature on the 'magic numbers' and overall cluster size-dependence of the enthalpies and Gibbs free energies of solid and molten Fe clusters is analyzed in Fig. 2 as a function of cluster size. Temperature was varied from 400 K (solid clusters) to 2200 K (liquid clusters). The entropy and Gibbs free energy were calculated at atmospheric pressure but can be recalculated for any pressure using Eq. (6).

At 400 K, cluster enthalpy per atom is a non-smooth and non-monotonic function of cluster size, *i* (Fig. 2a) with 'magic number' sizes (e.g., 13, 19 atoms, etc.) corresponding to local enthalpy minima. This effect originates from the dependence of cluster potential energy on geometric configuration of the clusters (i.e., proximity to a closed shell structure (Hoffenberg et al. 2024)). At higher temperatures, when clusters melt and their atoms move stochastically, geometric effects disappear and enthalpy becomes a smooth function of *i*. This happens for bigger clusters (*i* > 40 atoms) at first (at 1000 K), and then, at higher temperatures, for the entire cluster size range.

Entropy per atom, however, is a non-monotonic (and non-smooth) function of *i* for both solid and liquid clusters (Fig. 2b). Correspondingly, Gibbs free energy of cluster formation per atom – the sum of enthalpic and entropic contributions – is also non-monotonic in *i* at all temperatures (Fig. 2c). Note: the dips and spikes in the cluster free energy curves are considerably smaller than those for Al clusters (Girshick, Agarwal, and Truhlar 2009), potentially due to the absence of electronic effects in the Fe model.

The overall dependence of the Gibbs free energy on *i* is weaker than that of enthalpy, especially at higher temperatures. The enthalpy varies quite substantially from ~-150 kJ/mol for a dimer to < -300 kJ/mol for larger *i*. Surface effects of 'dangling' bonds on enthalpy are stronger for smaller clusters with a higher surface to volume ratio. Surface effects on entropy per atom are also strong. In the Gibbs free energy, these effects in enthalpy and entropy partially cancel out, especially at high temperatures where the contribution of the entropic term becomes larger (Eq. (3)).

The values of the Gibbs free energies were used to analytically calculate equilibrium cluster densities (Eq. (13)) for the conditions of sub-saturated vapor. Under these conditions, an equilibrium cluster distribution was established where cluster densities near-exponentially decrease with $i$. The analytical results were compared to the size distribution observed in direct MD simulation of Fe vapor condensation for the same conditions, showing good agreement (Appendix 4). These results validate the Gibbs free energy values and the method of their calculation.

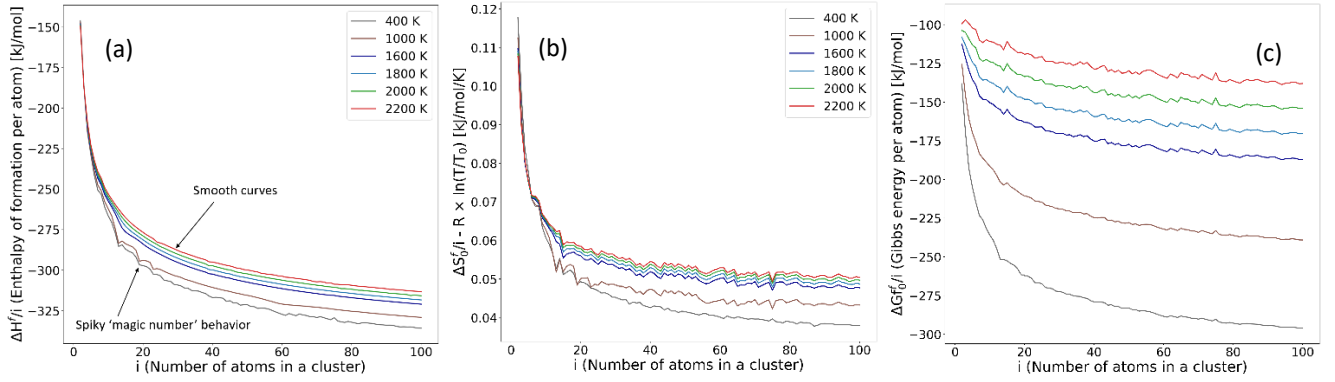

Fig. 2. Thermodynamic data of the clusters obtained in MD simulations: (a) enthalpy, (b) entropy, and (c) Gibbs free energy of cluster formation as function of cluster size for several temperatures at atmospheric pressure.

### 3.2. Effective surface tension coefficient and the Tolman correction

This section identifies and discusses an effective surface tension coefficient, $\sigma$, that results in accurate cluster Gibbs free energies when substituted into the spherical model. For this purpose, vapor pressure of the model Fe is identified first. Dependence on temperature and cluster size is analyzed and approximated by the Tolman correction. The effective surface tension coefficient is then compared to that obtained from cluster potential energies, which is representative of the $\sigma$ measured in oscillating droplet experiments (Bain 2024; Kasama et al. 1983; Klapczynski2022; Morohoshi et al. 2011; Ozawa et al. 2011; Wille, Millot, and Rifflet 2002) typically used for liquid metals. Finally, Gibbs free energy curves are plotted as a function of $i$ for various saturation degrees and compared to results using the spherical cluster approximation.

3.2.1. Vapor pressure of the model Fe

The effective surface tension coefficient (Eq. (10)) depends directly on the equilibrium vapor pressure, and indirectly on two material related constants, latent heat $L$, and boiling temperature at a reference pressure $T_{boil}$ through the Clausius-Clapeyron equation (9). For Eq. (10) to be self-consistent, the vapor pressure must be determined using the same Fe interaction potential used to obtain the cluster Gibbs free energies. In this regard, the parameters $L$ and $T_{boil}$ for the model Fe system can be regressed from the vapor pressure data obtained in direct coexistence MD simulations (Section 2.2).

Fig. 3 presents the MD simulation results for the logarithm of vapor pressure as a function of inverse temperature and comparison to experimental values. Using the Clapeyron-Clausius equation (9), the following values of the material parameters were regressed:

$$L = 4.21 \times 10^5 \, J/mol\,,$$

$$T_{boil} = 3125.8 \, K\,. \tag{14}$$

These values are close to the experimental data for Fe ($L_{exp} = 3.5 \times 10^5 \ J/mol$, $T_{boil,exp} = 3135 \ K$), however, the difference is large enough to affect the effective surface tension coefficient.

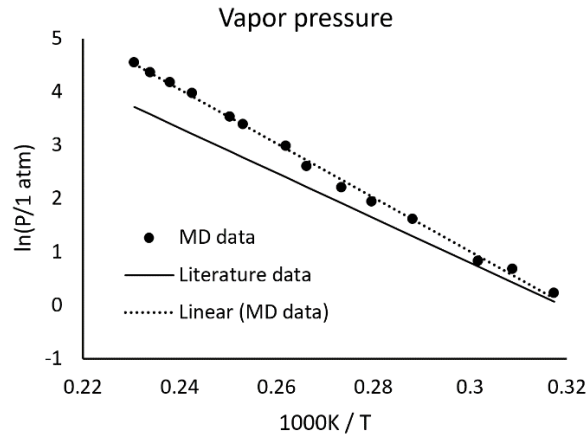

Fig. 3. Logarithm of the vapor pressure as a function of inverse temperature: results of direct vapor-liquid co-existence MD simulations. A linear fit and vapor pressure based on experimental values of the boiling point and latent heat of vaporization.

3.2.2. Effective surface tension coefficient and the Tolman correction

The effective surface tension coefficient (Eq. (10)) is plotted in Fig. 4 (thin solid lines) as a function of *i* for various temperatures. The effective surface tension generally increases with *i*, but exhibits high-frequency oscillations that reflect the non-monotonicity in Gibbs free energies (Fig. 1c). For the sake of analysis, it is desirable to separate the high-frequency oscillatory behavior from the increasing trend in the surface tension. The effect of these high-frequency oscillations on the vapor condensation and cluster growth process is addressed in Section 3.3. To describe the general dependence of surface tension on *i*, we can use the Tolman correction:

$$\sigma_i = \sigma \left(1 + \frac{2\delta_T}{r_W} i^{-1/3}\right)^{-1}, \qquad (15)$$

where $\sigma_i$ is the effective surface tension coefficient for a cluster of $i$ atoms, $\sigma$ is the surface tension coefficient of a flat surface, $\delta_T$ is the Tolman length, and $r_W$ is the Wigner-Zeitz radius of liquid Fe. The plots in Fig. 4 clearly suggest that in the model Fe system considered Tolman length is positive, resolving the debate on the sign of the Tolman length (Blokhuis and Kuipers 2006; Lei et al. 2005), at least for Fe.

The parameters $\sigma$ and $\delta_T$ in the Tolman expression (15) were determined from fitting the effective surface tension data (thin solid lines in Fig. 4) with expression (15) at each temperature. The fitting was done using the Scipy.Optimize Python module. Fig. 4 shows the fitted effective surface tension curves (thick solid lines) and the asymptotical values of the flat-surface $\sigma$ (horizontal dashed lines). The fitted parameters $\sigma$ and $\delta_T$ are presented in Fig. 5 as functions of temperature.

The Tolman correction provides a good fit for the general trend of the effective surface tension coefficient derived from the cluster Gibbs free energies. It only fails to describe the effective surface tension for the smallest clusters: dimers and trimers. The asymptotic flat-surface $\sigma$, however, is far off for clusters of fewer than 100 atoms, so the effective surface tension is expected to rise substantially for $i \gg 100$.

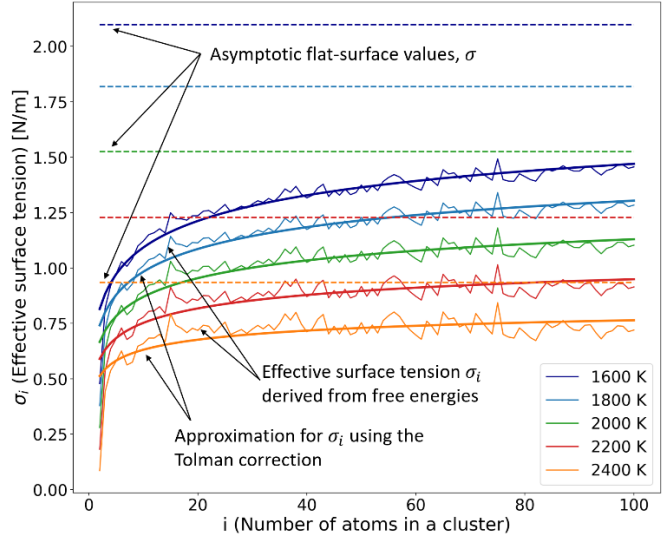

Fig. 4. Effective surface tension coefficient (thin lines) as a function of *i*, its approximation by the Tolman expression (15) (thick lines), and surface tension of a flat surface (dashed lines), all at various temperatures.

### 3.2.3. Dependence of surface tension and Tolman length on temperature

The asymptotic surface tension coefficient $\sigma$ corresponding to a flat surface, exhibits a strong dependence on temperature (black line in Fig. 5a). It can be fitted well by a linear dependence (gray dashed line) with the following slope $a_\sigma$ and intercept $b_\sigma$:

$$\sigma = a_\sigma T + b_\sigma, \ a_\sigma = -0.00146 \text{ N}/(m\ K), \ b_\sigma = 4.437 \text{ N/m} \tag{16}$$

The slope of this dependence is considerably higher than the one exhibited by experimental data for Fe.

It is handy to multiply the Tolman length, $\delta_T$, by $\sigma$, as the result can be approximated by a linear relation (Fig. 5b). The resulting slope $a_\delta$ and intercept $b_\delta$ are:

$$\frac{\delta_T}{r_W}\sigma = a_\delta T + b_\delta, \ a_\delta = -0.00205 \text{ N}/(m\ K), \ b_\delta = 5.345 \text{ N/m} \tag{17}$$

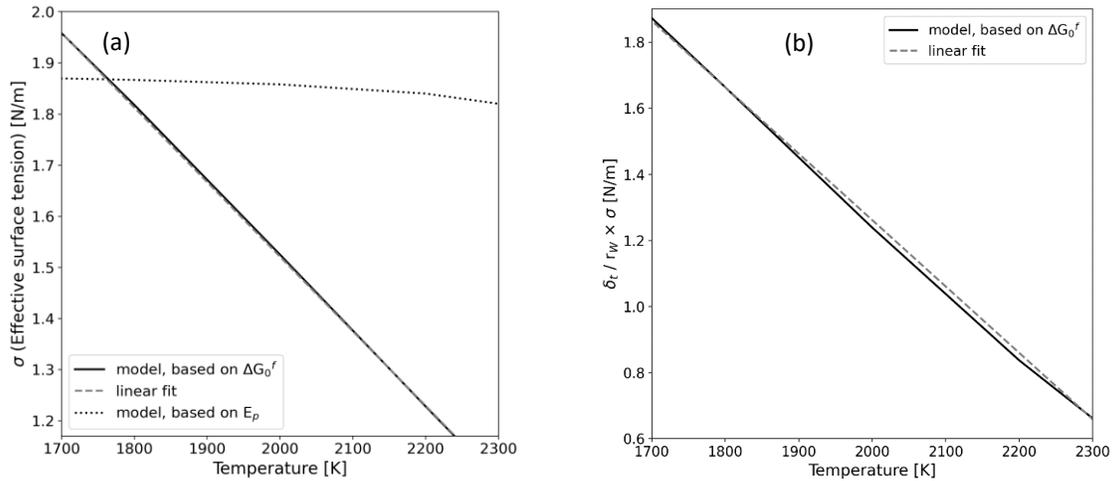

Fig. 5. (a) Surface tension coefficient for a flat surface, $\sigma$, and (b) the normalized Tolman length multiplied by the surface tension coefficient, $\sigma\delta_T/r_W$, as functions of temperature.

3.2.4. Deriving surface tension coefficient from potential energy

The classical nucleation theory (CNT) relies on using the flat-surface $\sigma$ for cluster Gibbs free energies (Eq. (7)). This approach ignores the dependence of the surface tension coefficient on *i* (important for small clusters), and relies on the *mechanical* (not thermodynamical) surface tension coefficient, defined by forces acting on a curved free surface. This surface tension coefficient is obtained in experiments with oscillating droplets (Klapczynski2022; Morohoshi et al. 2011; Ozawa et al. 2011; Wille, Millot, and Rifflet 2002) It represents the change of the droplet *potential* energy rather than its *free* energy in response to changes in droplet surface area.

Here, we aim to compare a surface tension coefficient extracted from cluster potential energies, $\sigma_p$, with the one using the cluster Gibbs free energies. Following CNT and assuming a constant surface tension coefficient, its value should be connected to the cluster potential energies through:

$$\frac{E_{p,i}}{i} = E_{p,bulk} + i^{-\frac{1}{3}} 4\pi N_A r_W^2 \sigma_p, \qquad (18)$$

Where subscript *p* in $\sigma_p$ denotes origination from cluster potential energies. $E_{p,bulk}$ is potential energy of bulk Fe liquid (per atom). Both $\sigma_p$ and $E_{p,bulk}$ are unknown functions of *T*.

For a given temperature, both parameters can be regressed from the curve $E_{p,i}/i$ as a function of *i*, for i > 20 atoms. The potential energy curves and their fits (which are very good) with Eq. (18) are presented in Fig. 6, for several temperatures.

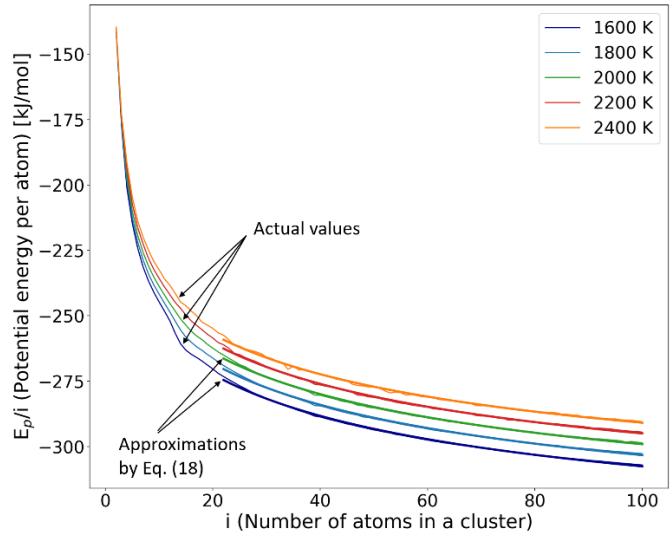

Fig. 6. Regression of the surface tension coefficient from cluster potential energy. Cluster potential energy per atom as a function of cluster size (thin lines) and fits with expression (18) (thick lines).

The potential-energy surface tension coefficient, $\sigma_p$, is in a good agreement with experimental data (both qualitatively and quantitatively (Fig. 7)) and exhibits weaker dependence on temperature than the thermodynamic $\sigma$ derived from cluster free energies (Fig. 5a). This result implies that: 1) the atomic interaction potentials for Fe accurately predict the potential energy of surface atoms and the effects of dangling bonds, and 2) the free energies calculated using the mechanical surface tension coefficient deviate from the actual ones. The difference between the mechanical and thermodynamic surface tensions is not surprising considering different behaviors of the cluster enthalpy and free energy (Figs. 2a,c).

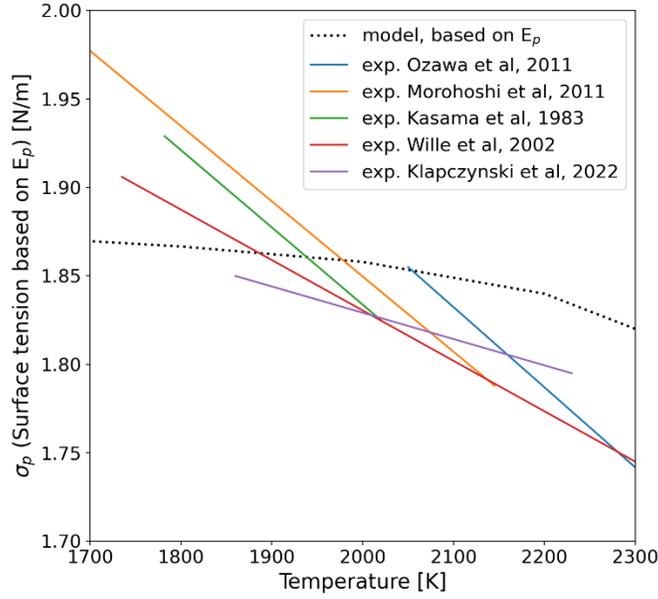

Fig. 7. Mechanical surface tension, $\sigma_p$, derived from cluster potential energies (black dotted line) and experimental data (color lines), as a function of temperature.

3.2.5. Gibbs free energy curves for supersaturated vapor

As discussed in Section 2, the Gibbs free energy of cluster formation depends on the vapor pressure (Eq. (6)) and on the saturation degree, $S$, (Eq. (7)). For a supersaturated vapor ($S > 1$), the free energy of cluster formation peaks at a curtain cluster size, referred to as the critical size ($i^*$). The free energy at this peak determines the cluster nucleation barrier. The higher the saturation degree, the smaller are the critical size and the nucleation barrier.

Gibbs free energy of cluster formation is plotted vs. $i$ for various saturation degrees, $S$, at a constant temperature of 2200 K (Fig. 8). The accurate Gibbs free energies, $\Delta G_i^f$ (calculated in Section 2.1) are compared to the spherical cluster approximation, $\Delta G_i^{f,spher}$ (Eq. (7)) with various surface tensions. Two *thermodynamic* surface tension coefficients – size-dependent coefficient $\sigma_i$ from the accurate Gibbs energies with the Tolman correction (Eq. (15)) and constant flat-surface $\sigma$ (the same Eq. (15)) – are compared to *mechanical* surface tension $\sigma_p$ regressed from cluster potential energies (Eq. (17), Fig. 7). Note: all variations of the spherical approximation produce smooth curves (high-frequency oscillation data was lost in the regressions).

The spherical approximation with the Tolman correction for surface tension, $\sigma_i$, (thick lines) agrees well with the accurate Gibbs free energies (thin solid lines) for all saturation degrees (if magic number oscillations are ignored). The other two variations of the spherical cluster approximation that rely on $\sigma$ and $\sigma_p$ (dashed and dotted lines, respectively), considerably overestimate the Gibbs energies, resulting in larger $i^*$ and higher nucleation barrier. The deviation is stronger for $\sigma_p$. At the temperature considered (2200 K), the value of $\sigma_p$ (dotted line in Fig. 7) is much higher than that of $\sigma$ (black solid line in Fig. 5a).

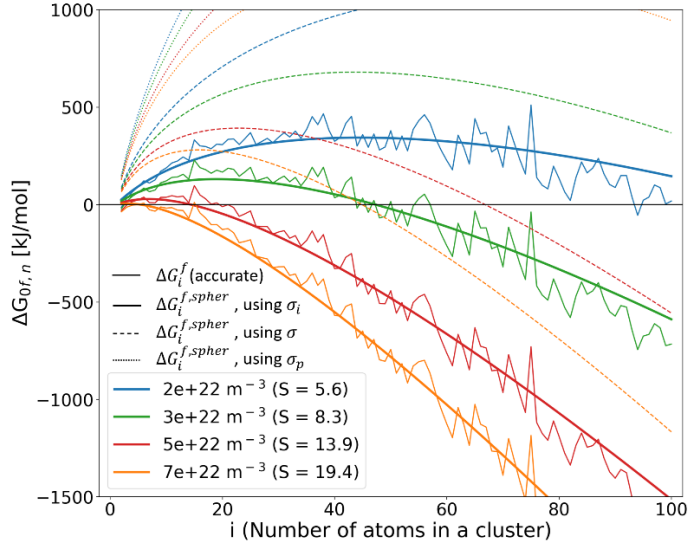

Fig. 8. Gibbs free energy of cluster formation vs. cluster size, *i*, for various vapor saturation degrees. Accurate free energies (Eq. (6)) (thin solid lines); spherical approximation with the Tolman correction, $\sigma_i$ (thick solid lines); spherical approximation without Tolman correction, $\sigma$ (dashed lines); spherical approximation based on the mechanical surface tension, $\sigma_p$ (dotted lines).

### 3.3. Modeling cluster formation and growth using accurate cluster free energies and spherical approximation

This section analyzes the use of various approximations for cluster free energies in the modeling of vapor condensation. The efficient kinetic solver for the particle balance equation (PBE; Eq. (11)) from Khrabry et al. (2024) was utilized. The results obtained using accurate Gibbs free energies (Section 3.1) are compared to those using the spherical approximation (Eq. (7)) with various surface tension coefficients ($\sigma_i, \sigma$, and $\sigma_p$, Section 3.2.5). The accurate Gibbs free energies were only used for clusters sizes up to 100 atoms; for bigger clusters, the spherical approximation with the Tolman correction was used. A general description of the condensation process is given first followed by a discussion of differences from various approximations for the cluster free energies.

Condensation of Fe vapor was modeled for two cooling rates of $10^5$ K/s and $10^6$ K/s, representative of fast vapor cooling in a Laval nozzle (Rao et al. 1995; Zhalehrajabi and Rahmanian 2014), spark discharge (Maisser et al.2015), pulse laser ablation (Wang et al. 2012), and on the high end of those reported in quenched thermal plasma streams (Shigeta, Hirayama and Ghedini 2021). Such rapid vapor cooling is required to efficiently produce high quantities of Fe nanoparticles with small diameters of one to several nanometers, which are in demand for efficient synthesis of high-quality, few-walled carbon nanotubes with minimal iron consumption for this matter. The diameter of nanoparticles produced is roughly proportional to the vapor density, and inversely proportional to the cooling rate (Tacu, Khrabry, and Kaganovich 2020).

The modeled Fe vapor is condensed from initially saturated conditions (*S*=1) at 1800°C. The modeling is performed until the temperature drops to 900°C, which is typical for the chemical-vapor-deposition (CVD) growth of carbon nanotubes from hydrocarbon gaseous precursors (Hoecker et al. 2016; Puretzky et al. 2005). This corresponds to 9 ms and 0.9 ms in the cases of $10^5$ K/s and $10^6$ K/s cooling rates, respectively.

The kinetic solver from Khrabry et al. (2024) models the evolution of the entire cluster size distribution starting from dimers and accounting for all processes associated with cluster growth. These include cluster nucleation, surface growth through condensation, cluster evaporation, and collisions leading to clusters merging (coagulation/coalescence) into larger clusters.

Condensation does not happen immediately as the vapor cools down from the saturation point (Fig. 9a, Fig. 11a). Monomer density stays virtually constant (or very sightly decreasing) for the first couple of nanoseconds or tens of a nanosecond, depending on the cooling rate. Initially, when the saturation degree (Figs. 9b and 11b) exceeds 1 only slightly, *i\** (Figs. 9 and 11c) and the nucleation barrier are large, causing the delay. As the saturation degree increases, the nucleation barrier and *i\** both decrease until the onset of the condensation occurs. Clusters start forming and the vapor further condenses on them, leading to a drop in monomer density and an increase in both cluster density (Figs. 9d and 11d) and average size (Figs. 9e and 11e). This rapid increase in the cluster density is commonly referred to as a *nucleation burst* (Warren and Seinfeld 1984). The analytical solution by Tacu, Khrabry, and Kaganovich (2020) provides expressions for the cluster condensation time and average diameter of forming clusters, assuming a constant surface tension coefficient in the spherical model.

Only clusters that are bigger than the critical size – super-critical clusters – are accounted for in Figs. 9d-f and 11d-f. Clusters below the critical size – sub-critical clusters – are thermodynamically unstable, that's why most other codes and models (Friedlander 1983; Frenklach and Harris 1987; Frenklach 2002; Mitrakos, Hinis, and Housiadas 2007; Nemchinsky and Shigeta 2012; Prakash, Bapat, and Zachariah 2003), omit clusters below the critical size. In those models, clusters *appear* already having the critical number of atoms, at a rate given by an analytical expression from Girshick and Chiu (1990), assuming a spherical approximation with a constant surface tension coefficient. In the model by Khrabry et al. (2024) used in this work, we directly resolve sub-critical clusters because they actually play an important role in the formation of super-critical clusters, but only include super-critical clusters in the analysis of average cluster characteristics (in Figs. 9d-f and 11d-f).

After the nucleation burst takes place, the cluster density declines for two reasons. First, as the vapor condenses, the saturation degree decreases, increasing the critical cluster size and leading to a reduction in the number of super-critical clusters. Second, the smaller clusters merge into bigger ones through collisions (coagulation/coalescence).

There are many differences in the results between various approximations of the cluster free energies. These include effects on the timelines of cluster formation, cluster densities, and average diameters. The spherical approximation with constant surface tension coefficients $\sigma$ and $\sigma_p$ (Eq. (7)) yields significantly delayed condensation (for both $10^5$ K/s and $10^6$ K/s cooling) compared to accurate free energies or the size-dependent surface tension $\sigma_i$ with the Tolman correction.

Notably, when the free-energy-based constant surface tension $\sigma$ is used, the condensation does not initiate at all within the timeframes modeled. The density of monomers remains unchanged, and cluster density stays virtually at zero. The delay in the nucleation burst is due to much higher energy barriers for cluster formation when constant coefficients $\sigma$ and $\sigma_p$ are used (Section 3.2.5, Fig. 8). When the clusters finally formed in the $\sigma_p$ case, their diameter was about twice as small, and the cluster number density was ~two times higher than that modeled with the accurate cluster free energies.

The difference between the results for the accurate Gibbs free energies and those derived from $\sigma_i$, however, is small. These results suggest that the high-frequency magic-number oscillations in cluster free energies (Fig. 8) have a minor effect on how the Fe condensation process transpires. For both cooling rates, the solutions obtained using accurate free energies and spherical cluster approximation with the Tolman correction, $\sigma_i$, which was derived to fit these original free energies (thin and thick solid lines, respectively) are close in the evolution of vapor density (Figs. 9a and 11a), corresponding saturation degree (Figs. 9b and 11b), cluster density (Figs. 9d and 11d), average cluster size (Figs. 9e and 11e) and its standard deviation (Figs. 9f and 11f) (i.e., the width of the cluster size distribution).

The differences between the accurate cluster free energy and the $\sigma_i$ solutions, while minor, are:
1) Monomer density immediately drops by tens of percent (Figs. 9a and 11a) with accurate free energies, which is not observed with $\sigma_i$. This drop is attributed to higher densities of dimers in the solution with accurate free energies (cf. Figs.10a,b, and Figs. 12a,b). Dimers have a very low free energy of formation which cannot be captured by the spherical approximation (Section 3.2.2, Fig. 4), even with the Tolman correction. Hence, in the solution with accurate cluster energies, dimers have a higher density consuming a notable fraction of vapor atoms. This result points at the importance of including the entire cluster size distribution in the model (starting from dimers, not just super-critical clusters).
2) Lower monomer densities in the accurate cluster energy case imply a proportionally lower saturation degree (Fig. 9b). However, the effect of the lower saturation degree on nucleation burst timing and the cluster formation process as a whole is small.
3) The critical cluster size experiences abrupt changes with time (Figs. 9c and 11c) in the solution with accurate cluster free energies. Since the accurate energy curve is non-monotonic, the critical cluster size *jumps* from one local peak in the energy curve to a neighbor one as the curve evolves with the change in saturation degree. These jumps can only be modeled for cluster sizes under 100 atoms, for which accurate free energies were calculated. In the case of slow cooling (Fig. 9c), $i^*$ only dips into this range for a short time, then bounces back to higher values. In the faster cooling case, however, the critical size stays below 100 atoms for most of the condensation process, eventually decreasing in a stepwise fashion to just several atoms (Fig. 11c).
4) The non-monotonicity of accurate free energies is reflected and amplified in the cluster size distribution (Figs. 10a and 12a): spikes and troughs with a difference of orders of magnitude are observed in densities of clusters of 2-100 atoms. This result implies that only certain cluster sizes should be observed during condensation and others should have negligible densities.

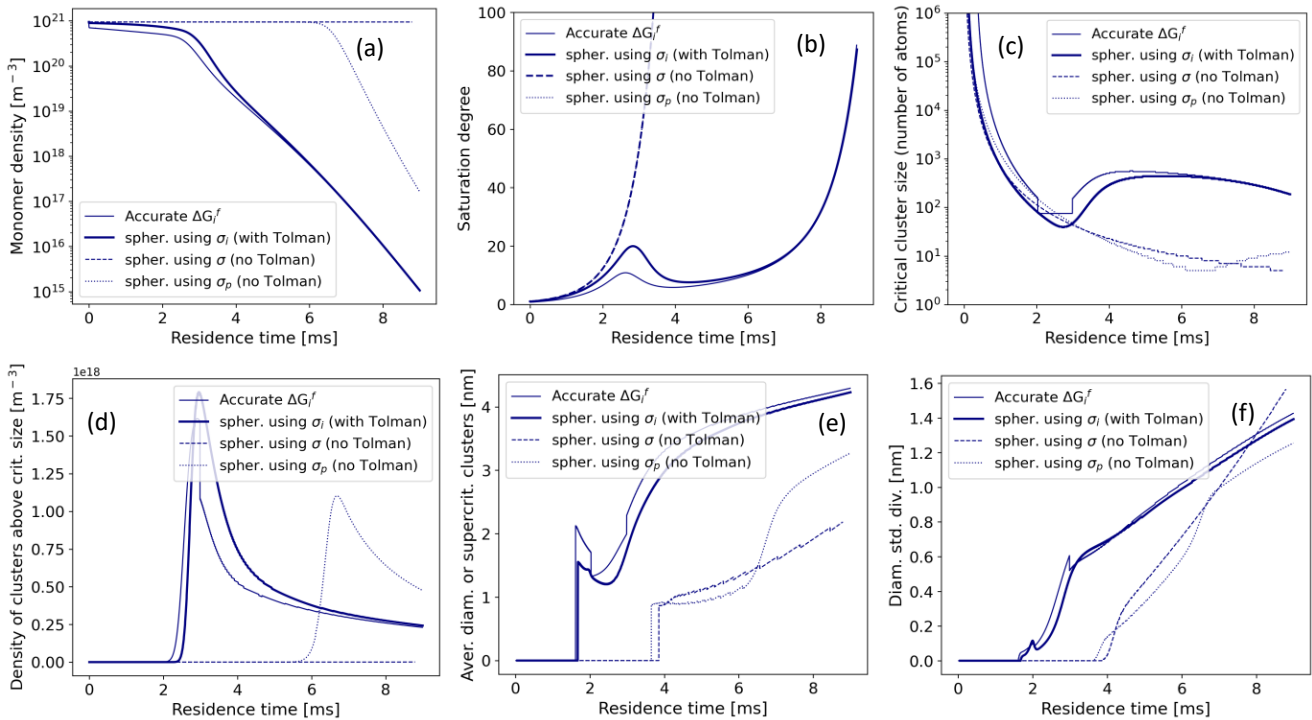

Fig. 9. Time evolution of the Fe vapor and particle distribution parameters during condensation of initially saturated vapor cooling down from 1800°C at $10^5$ K/s. Modeling results using accurate Gibbs free energies (Section 3.1; thin solid lines), and the spherical cluster approximation (Eq. (7)) with various surface tension coefficients: size-dependent surface tension with Tolman-correction $\sigma_i$ (thick solid lines), constant flat-surface $\sigma$ (dashed lines), and constant surface tension $\sigma_p$ derived from cluster potential energies (dotted lines).

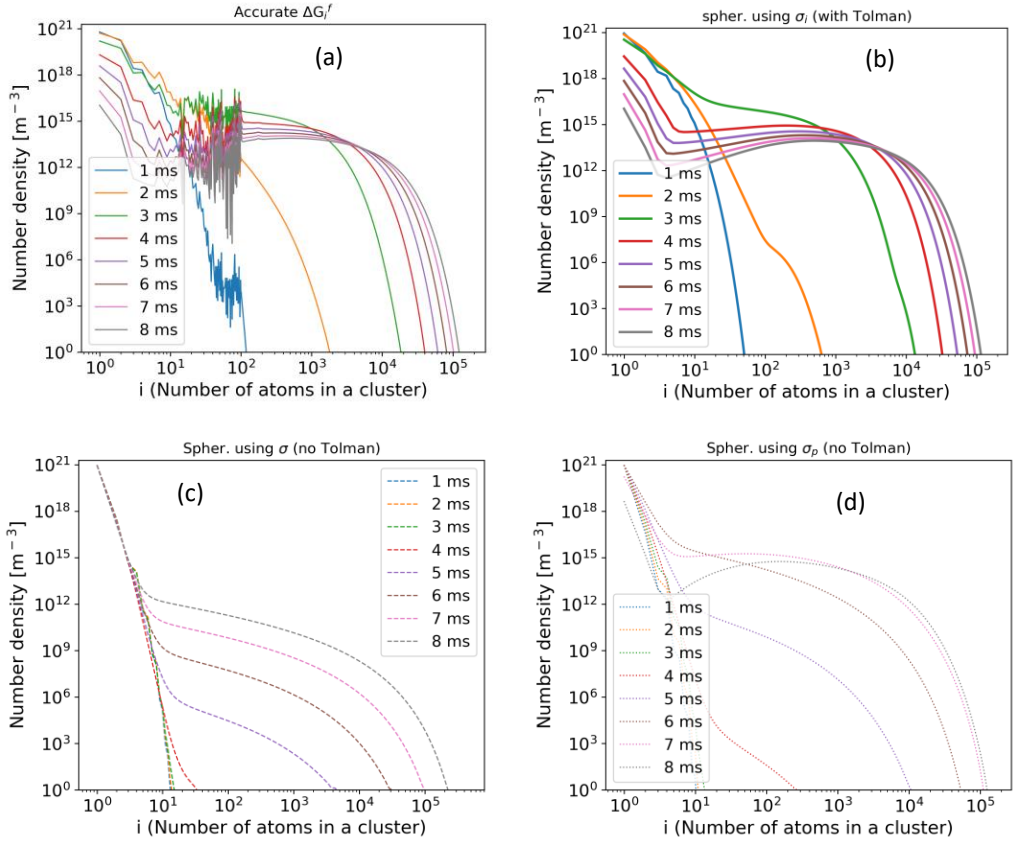

Fig. 10. Time evolution of the cluster size distribution during condensation of initially saturated Fe vapor cooling down from 1800°C at $10^5$ K/s. Modeling results using (a) accurate Gibbs free energies (Section 3.1), and the spherical cluster approximation (Eq. (7)) with various surface tension coefficients: (b) size-dependent surface tension with Tolman correction $\sigma_i$, (c) constant flat-surface $\sigma$, and (d) constant surface tension $\sigma_p$ derived from cluster potential energies.

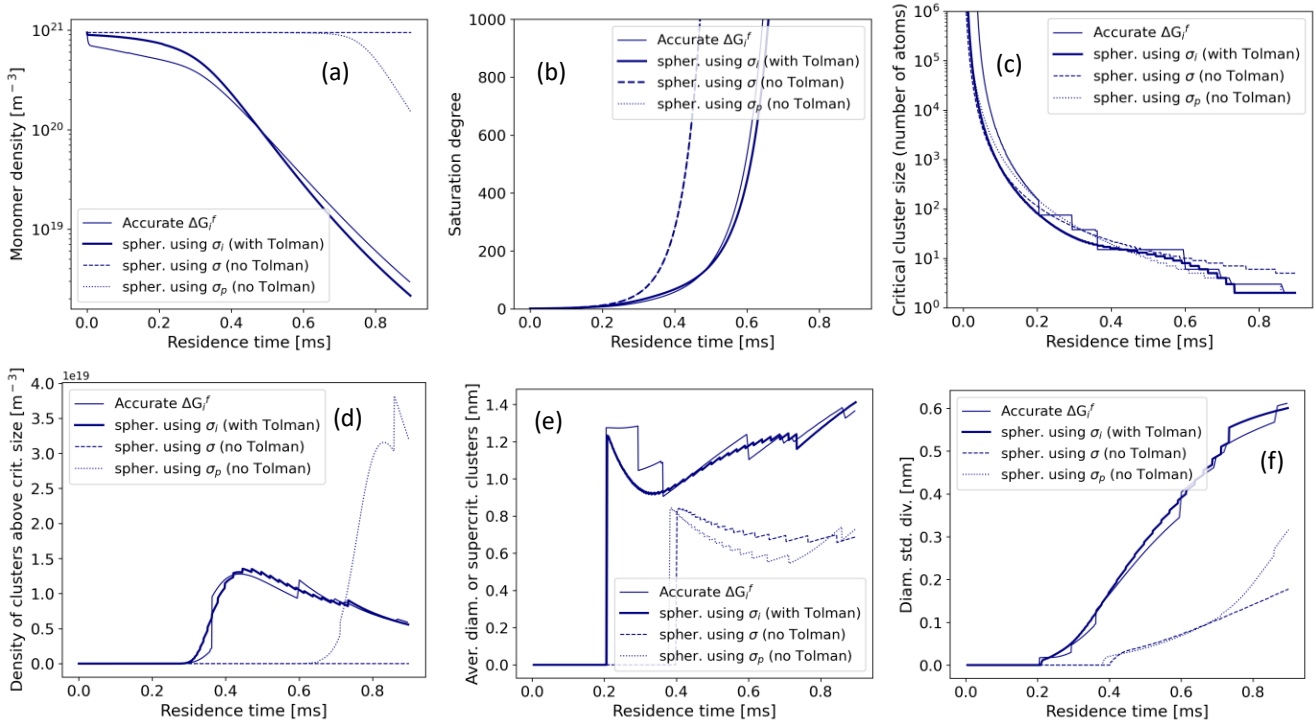

Fig. 11. Time evolution of the Fe vapor and particle distribution parameters during condensation of initially saturated vapor cooling down from 1800°C at $10^6$ K/s. Modeling results using accurate Gibbs free energies (Section 3.1; thin solid lines), and the spherical cluster approximation (Eq. (7)) with various surface tension coefficients: size-dependent surface tension with Tolman-correction $\sigma_i$ (thick solid lines), constant flat-surface $\sigma$ (dashed lines), and constant surface tension $\sigma_p$ derived from cluster potential energies (dotted lines).

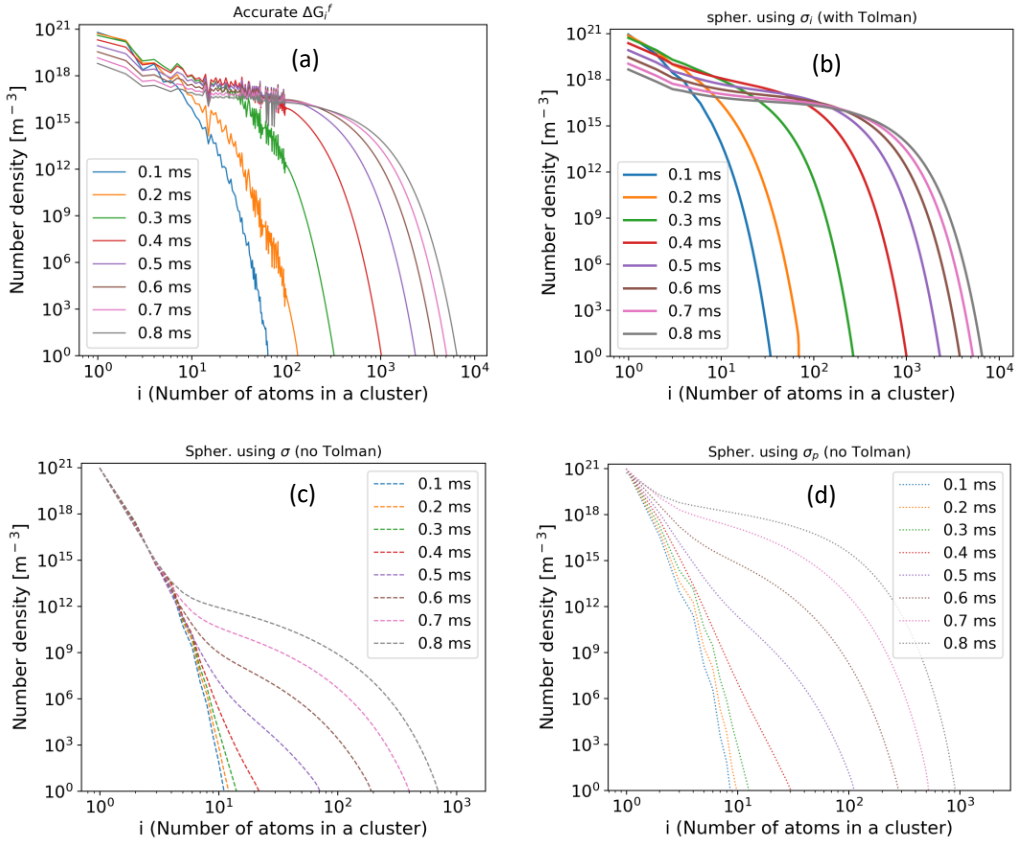

Fig. 12. Time evolution of the cluster size distribution during condensation of initially saturated Fe vapor cooling down from 1800°C at $10^6$ K/s. Modeling results using (a) accurate Gibbs free energies (Section 3.1), and the spherical cluster approximation (Eq. (7)) with various surface tension coefficients: (b) size-dependent surface tension with Tolman correction $\sigma_i$, (c) constant flat-surface $\sigma$, and (d) constant surface tension $\sigma_p$ derived from cluster potential energies.

### 3.4. Analytical expression for cluster nucleation rate accounting for the size-dependence of clusters' surface tension coefficient

Detailed kinetic modeling of the cluster size distribution discussed in Section 3.3 showed that the spherical approximation for Gibbs free energies yields reasonable cluster formation dynamics if amended with the Tolman correction. However, commonly used models of cluster formation and growth (Friedlander 1983; Panda and Pratsinis 1995; Prakash, Bapat, and Zachariah 2003; Warren and Seinfeld 1985) don't directly resolve sub-critical clusters. Instead, these models rely on an analytical expression for the cluster nucleation rate (A3.19) derived by Girshick and Chiu (1990), which assumes constant surface tension coefficient. This assumption leads to inaccurate modeling results for Fe clusters (Section 3.3), so this expression needs to be expanded to account for the size-dependent surface-tension coefficient given by the Tolman correction. Rao and McMurry (1990) derived a Tolman-corrected expression for the nucleation rate. However, they used an non-self-consistent for the cluster's free energy which predicts non-zero energy of monomer formation from itself.

In Appendix 3, a new expression for the nucleation rate is derived, incorporating the Tolman correction and relying on a self-consistent cluster free energy expression (7). This new expression (Eq. (A3.18)) is essentially an expansion of Girshick's original expression, and reduces to the original if the Tolman length is set to zero (making the surface tension coefficient a constant).

Fig. 13 compares the nucleation rate predicted by the full numerical model (including clusters of all sizes) to the rate given by the analytical expression (A3.18). The vapor saturation degree, $S$, in Eq. (A3.18) was taken from the numerical solution for consistency. Two variations of the numerical model were used, one including cluster merging (coagulation; Fig. 13a), and one omitting coagulation (Fig. 13b) by ignoring the coagulation term in Eq. (11). Omitting coagulation makes the numerical model assume atom-wise condensation and evaporation, similarly to the assumptions underlying the analytical expression (A3.18).

Upon omission of coagulation in the numerical model (Fig. 13b), the nucleation rate predicted by the analytical expression agrees well with the numerical results. When the coagulation is included in the numerical solution (Fig. 13a), the analytical expression underestimates the nucleation rate by orders of magnitude. This disagreement indicates that coagulation of sub-critical Fe clusters plays a crucial role in the nucleation process and drastically accelerates the rate at which the clusters surpass the nucleation barrier. This effect is prominent for Fe clusters because of retained high density of sub-critical clusters (Figs. 10a,b and Figs. 12a,b), so that collisions between sub-critical clusters cannot be ignored. This result indicates that the inclusion of both coagulation and subcritical clusters in the nucleation model of Fe condensation is necessary.

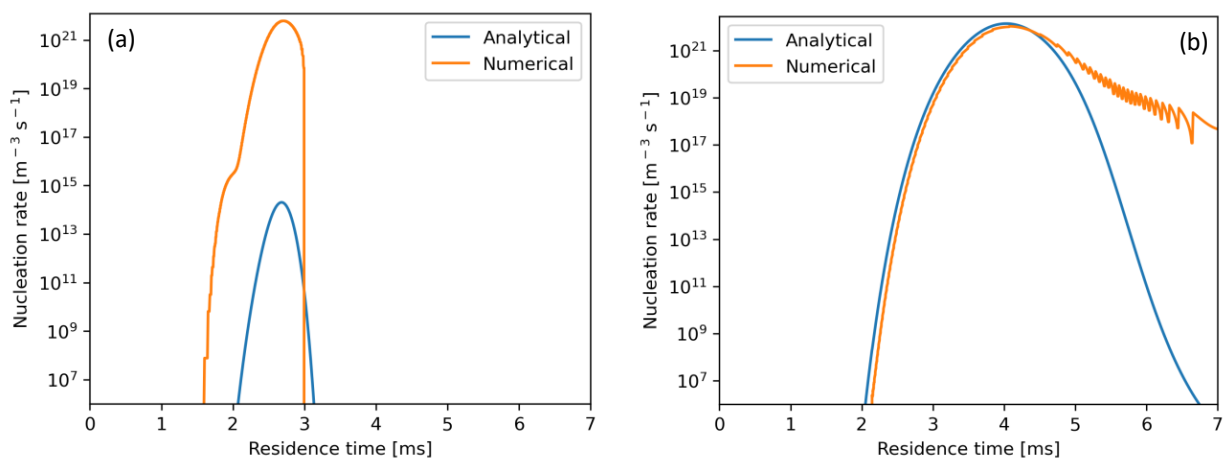

Fig. 13. Time evolution of the Fe cluster nucleation rate during condensation of initially saturated vapor cooling down from 1800°C at $10^5$ K/s. Spherical approximation with the Tolman correction is used for the cluster free energies. Cluster coagulation is (a) included, and (b) omitted.

## 4. Summary

A simple method of calculating Gibbs free energies of cluster formation from MD simulation data has been developed. These data include: 1) time-average kinetic and potential energies of individually simulated clusters, and 2) the Hessian (or second derivative) matrix of the energy with respect to atomic displacements.

The method has been applied to calculate Gibbs free energies of formation of small Fe clusters with 2 to 100 atoms. The classical MD simulation data for individual Fe clusters was taken from Hoffenberg et al. (2024), which used a classical many-body interatomic potential for Fe.

The free energies of Fe clusters were verified by applying them to calculate equilibrium cluster size distributions in sub-saturated vapors and comparing them to those directly observed in the MD modeling for the same conditions using the same classical inter-atomic potentials. Good agreement was observed.

An effective size-dependent surface tension coefficient, $\sigma_i$, for Fe clusters up to 100 atoms was calculated to yield accurate free energies for each cluster size if substituted in the spherical cluster model. This effective surface tension depends strongly on both the cluster size and temperature. The dependence on the cluster-size is non-monotonic, with high-frequency peaks and troughs owed to magic number clusters. The Tolman correction accurately described the general cluster size dependence of the effective surface tension (without high-frequency magic number oscillations), except at the smallest cluster sizes of 2-3 atoms.

A surface tension coefficient for a flat surface $\sigma$ was regressed using the Tolman correction. It was larger than the effective coefficients for the small clusters of 2-100 atoms $\sigma_i$. The flat-surface coefficient $\sigma$ yields over-estimated critical cluster sizes and the cluster nucleation energy barrier.

Additionally, the *mechanical* surface tension, $\sigma_p$, was derived from the cluster potential energies. Its values appeared qualitatively and quantitatively different from the *thermodynamic* coefficient $\sigma$ derived from cluster free energies. However, $\sigma_p$ appeared to be in a considerably better agreement with the available experimental data, implying that: 1) the classical MD inter-atomic potentials for Fe accurately predict potential energy of the surface atoms and the effects of dangling bonds, and 2) the free energies calculated using $\sigma_p$ (which is measured in the experiments) deviate substantially from the actual ones.

Cluster size distributions during vapor condensation were calculated with an efficient kinetic solver using various approximations for the cluster free energies. The solutions obtained using accurate cluster free energies were compared to those obtained using the spherical cluster approximation with various surface tension coefficients: $\sigma_i$ (Tolman corrected), $\sigma$ and $\sigma_p$ (not size-dependent).

The flat-surface surface tensions, $\sigma$ and $\sigma_p$, both yielded delayed condensation compared to the use of Tolman-corrected $\sigma_i$ or accurate free energies. The delay is attributed to much higher energy barriers for cluster formation. The $\sigma$ and $\sigma_p$ cases also yielded smaller diameter clusters with higher number density compared to $\sigma_i$ and accurate free energies.

The modeling results are quite similar between the cases of the Tolman corrected, *thermodynamic* $\sigma_i$ and the accurate Gibbs free energies. The minor differences between these cases are:

1. The non-monotonic spikes of the accurate free energies (from *magic number* effects) is reflected and amplified in the cluster size distribution, while the $\sigma_i$ case yields a smooth distribution. This result implies that only certain cluster sizes should be observed during condensation and others should have negligible densities. Despite this qualitative difference, the overall shape of the cluster size distribution is close in these two solutions.
2. The critical cluster size, *i\**, experiences abrupt changes when accurate free energies are used. Since the energy curve is non-monotonic, the critical cluster size *jumps* from one local maximum to a neighbor one as the curve evolves due to changing saturation degree.

However, despite these qualitative differences between the accurate free energies and the spherical model with the Tolman correction, the overall evolution of the cluster size average and the distribution width are quite close.

The authors hope that these findings will enhance nucleation modeling and enable better control over Fe nanoparticle and carbon nanotube synthesis.

## Appendix 1. Calculating entropy of formation of a solid cluster

The entropy of a cluster formation is, by definition, a difference between its entropy and entropies of $n$ gas-phase atoms that are required to form it. The entropy of a solid cluster is a sum of entropies of its internal (i.e., vibrational), translational and rotational motion. This yields the following expression for the cluster's energy of formation at atmospheric pressure:

$$\Delta S_0^f(T_0) = S_{vib}(T_0) + S_{tra,0}(T_0) + S_{rot}(T_0) - iS_1. \tag{A1.1}$$

Here, $S_{vib}$, $S_{tra}$ and $S_{rot}$ are vibrational, translational, and rotational entropies of the *i*-atomic cluster, and $S_1$ is translational entropy of an atom. They are defined as follows (McQuarrie and J. D. Simon 1999; Ochterski 2000). The translational entropy is given by:

$$S_{tra,0}(T_0) = R\left(\ln\left(\left(\frac{2\pi imkT_0}{h^2}\right)^{3/2}\frac{kT_0}{p_0}\right) + \frac{5}{2}\right), \tag{A1.2}$$

where $m$ is the mass of an atom, $h$ is the Planck constant, $p_0$ is atmospheric pressure, $i$ is the number of atoms in a cluster ($i = 1$ in $S_1$).

The vibrational entropy is given by:

$$S_{vib}(T_0) = R\sum_j\left(\frac{\Theta_{v,j}}{e^{\Theta_{v,j}}-1} - \ln(1-e^{-\Theta_{v,j}})\right), \tag{A1.3}$$

where $\Theta_{v,j}$ is the non-dimensional characteristic vibrational temperatures defined as:

$$\Theta_{v,i} = \frac{h\nu_j}{kT_0}. \tag{A1.4}$$

Here, $\nu_j$ is a vibrational frequency in rad/s. The summation in (A1.3) is over all vibrational frequencies of a cluster. A solid cluster has 3*i*-6 vibrational frequencies, where *i*>2 is the number of cluster's atoms. These vibrational frequencies were defined from the Hessian, or second derivative, matrix for the cluster energy with respect to atom displacements, in the absolute minimum cluster energy configuration (when the first derivatives of the energy with respect to displacement of the atoms are zero), as is described in detail in (Ochterski 1999, 2000). The Hessian matrix was determined in the MD modeling code LAMMPS (Thompson2022) by applying tiny displacements in each of three coordinate directions to individual atoms of a cluster in an optimized (minimum energy) configuration. Then the matrix was normalized by the atom mass, transformed to internal cluster coordinates to exclude the effects of rotational motion, and diagonalized using a Python module Sympy to identify its eigenvalues $\lambda_i$ corresponding to the vibrational motion in the cluster. The vibrational frequencies are simly defined as $\sqrt{\lambda_i}/(2\pi)$.

The rotational entropy of a dimer is given by:

$$S_{rot}(T_0) = R(-\ln\sigma_r - \ln\Theta_r + 1), \tag{A1.5}$$

where $\sigma_r$ is the rotational symmetry number of the cluster, $\Theta_r$ is the non-dimensional characteristic rotational temperature of the cluster defined as:

$$\Theta_r = \frac{h^2}{8\pi^2 kIT_0}. \quad (A1.6)$$

Here, $I$ is main moment of inertia of the cluster.

The rotational entropy of bigger clusters is given by:

$$S_{rot} = R\left(\ln\frac{\pi^{1/2}}{\sigma_r} - \frac{1}{2}ln(\Theta_{r,x}\Theta_{r,y}\Theta_{r,z}) + \frac{3}{2}\right), \quad (A1.7)$$

where $\Theta_{r,x}$, $\Theta_{r,y}$ and $\Theta_{r,z}$ are non-dimensional characteristic rotational temperatures defined as:

$$\Theta_{r,x} = \frac{h^2}{8\pi^2 kI_{xx}T_0}, \quad \Theta_{r,y} = \frac{h^2}{8\pi^2 kI_{yy}T_0}, \quad \Theta_{r,z} = \frac{h^2}{8\pi^2 kI_{zz}T_0} \quad (A1.8)$$

Here, $I_{xx}$, $I_{yy}$ and $I_{zz}$ are cluster's moments of inertia with respect to its three principal axes.

Note that the relations (A1.1)-(A1.8) have been derived from quantum statistical mechanics (McQuarrie and Simon 1999). They account for a discrete nature of energy levels in the cluster, which manifests in the involvement of the Planck constant. These relations are accurate for any temperature, including temperatures below the energy quant $hv_j/k$, given that the cluster behaves as a harmonic oscillator. However, the classical definition of temperature, Eq. (1), is only valid for higher temperatures.

One way to resolve this seeming contradiction would be to use the initial temperature $T_0$ (the lower bound in the integral (5)) above the rotational and energy quants in order to make the Eq. (1) relevant. This corresponds to temperatures above approximately 500K in the Fe clusters considered. Unfortunately, this temperate is high enough to cause non-harmonicity in the cluster's oscillations (the response force to atoms' displacements becomes substantially non-linear) rendering Eq. (A1.3) inaccurate.

A more appropriate way is to look at this problem from another angle. We are interested in cluster Gibbs free energies at temperatures of cluster condensation and growth, i.e., 1500K – 2300K. At these temperatures, quantum effects are not important. This implies that the results should not depend on the value of the Planck constant. In other words, the Planck constant can be assigned an arbitrary value, and the result mast still be the same, given that the energy quant is still below the temperature of the cluster. This logic justifies the use of MD modeling which does not incorporate the Plank constant in any way. For convenience, consider the a very low value of the Planck constant. In this case, Eq. (1) becomes justified at low temperatures, and Eq. (A1.3) simplifies to:

$$S_{vib} = R\sum_j(1 - ln\Theta_{v,j}), \quad (A1.9)$$

Substituting Eqs. (A1.2), (A1.7) and (A1.9) into Eq. (1.1) yields the following relation for the cluster's entropy of formation:

$$\Delta S_0^f(T_0) = R\left[\sum_j(1 - ln\theta_{v,j}) + \left(\ln\frac{\pi^{1/2}}{\sigma_r} - \frac{1}{2}ln(\theta_{r,x}\theta_{r,y}\theta_{r,z}) + \frac{3}{2}\right) + \frac{3}{2}\ln i + (1 - i)\left(\ln\left((2\pi m)^{3/2}\frac{(kT_0)^{5/2}}{p_0}\right) + \frac{5}{2}\right)\right], \quad (A1.10)$$

where

$$\theta_{r,x} = \frac{1}{8\pi^2 kI_{xx}T_0}, \quad \theta_{v,j} = \frac{v_j}{kT_0}.$$

Note that the Planck constant got cancelled out in Eq. (A1.10) which was used to determine initial entropy of cluster formation in the present study. This is consistent with the fact that the results don't depend on the Plack constant.

As is evident from Eqs. (A1.1)- (A1.5) and Eq. (A1.10), the only term in the definition of the entropy of formation (A1.1) that depends on pressure is $S_{tra}$ defined in Eq. (1.2). Thereby, if the entropy of a cluster formation $\Delta S_0^f(T_0)$ is known at atmospheric pressure, the entropy value $\Delta S^f(T_0)$ for an arbitrary pressure $p$ can be obtained from a simple relation:

$$\Delta S^f(T_0) = \Delta S_0^f(T_0) - (i-1)\ln\frac{p}{p_0} \qquad (A1.11)$$

**Appendix 2. Benchmarking the classical inter-atomic potential by comparing to DFT**

We performed DFT modeling of Fe clusters to benchmark the Finnis-Sinclair classical inter-atomic potentials that we use in MD modeling. 1) We obtained a caloric curve for a selected cluster size, $Fe_{13}$, and compared it to the one from the classical MD (Fig. A1a). 2) We calculated binding energies for various cluster sizes and compared them to those obtained using the classical potential (Fig. A1b).

To obtain the caloric curve of $Fe_{13}$ cluster, ab-initio MD (AIMD) modeling was performed using PBE DFT functional (Ernzerhof and Perdew 1998) in the VASP software package (Kresse and Hafner 1993). The conditions in the AIMD modeling were the same as in the classical MD. Each datapoint on the plot (Fig. A1a) corresponds to an individual NVE simulation, where atom movement is initialized with random velocities constrained in such a way that the cluster as a whole has no translational or rotational motion. The AIMD results appear considerably noisier because of the limitations on the time modeled from the computational cost. Each AIMD simulation was run for 1 ps, compare to 10 ns in each MD run. The classical MD curve is smooth, with very little noise, showing that the statistical errors in the time-averaged MD results are low. Aside from this different, the overall agreement between MD and AIMD results is quite good. The slope of the caloric curve is the same, and the inflection in the curve representing luster melting is a roughly the same temperature.

Binding energies of the clusters in a ground state (0K temperature) were calculated using two DFT functionals: the HSE06 functional (Henderson et al. 2009; Heyd and Scuseria 2004a, 2004b; Heyd et al. 2005; Heyd, Scuseria, and Ernzerhof 2006; Izmaylov, Scuseria, and Frisch 2006; Krukau et al.2006) implemented in the Gaussian-16 code (Frisch et al. 2016) and the PBE functional implemented in the VASP and Gaussian-16 codes. The results of the two codes (VASP and Gaussian-16) calculated using the same functional (PBE) are very close (Fig. A1b), demonstrating that the results don't depend on implementation in a code. Interestingly, the difference between the results obtained using two DFT functionals is considerably bigger that the different between the results of the classical (Finnis-Sinclair) potentials (Finnis and Sinclair 1984) and the PBE DFT functional. In other words, the uncertainty in the cluster energies predicted using DFT modeling is bigger that a potential error from using the classical potential. This result suggests that the classical (Finnis-Sinclair) potential can be employed to calculate cluster energies for a caloric curve, without increasing the uncertainty of the results compared to DFT modeling.

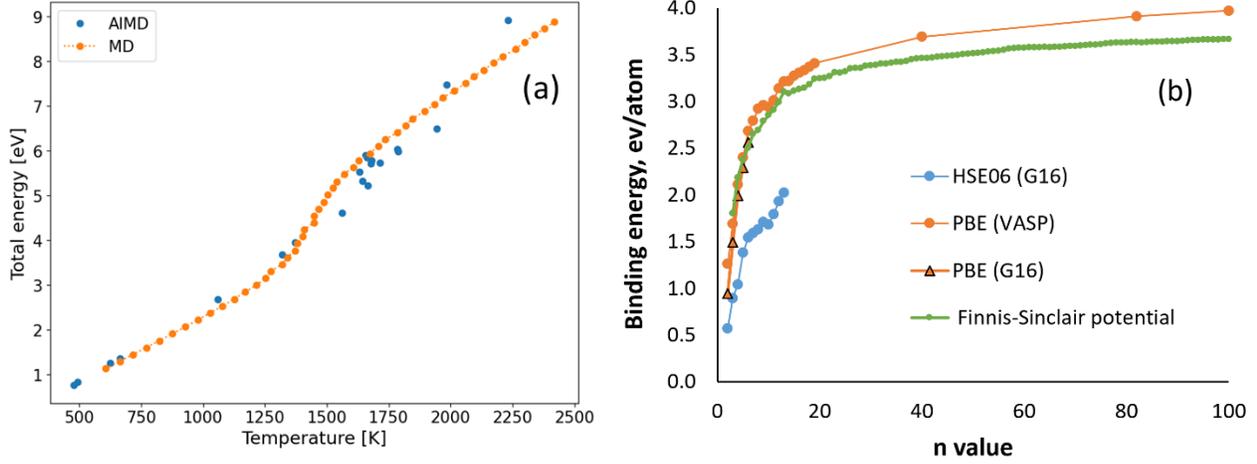

Fig. A1. Caloric curve for the Fe$_{13}$ cluster: total energy of the cluster (kinetic + potential energy) as a function of temperature (a); the binding energy per atom of Fe$_n$ clusters calculated by the HSE06, PBE DFT functionals and by the Finnis-Sinclair potential (b)

**Appendix 3. Deriving a nucleation rate expression for clusters with the Tolman correction**

As has been shown in the classical nucleation theory (CNT; Girshick and Chiu 1990), the rate of creation of super-critical clusters, commonly referred in literature as the cluster nucleation rate , can be expressed through the inverse sum of equilibrium cluster densities for all cluster sizes $i$.

$$J = \frac{v_{th}}{4} n_1 \left( \sum_{i=1}^{\infty} \frac{1}{s_i n_i^e S^i} \right)^{-1}. \tag{A3.1}$$

This expression was derived using the quasi-equilibrium assumption for sub-critical clusters, i.e., the rate of cluster creation is equal to the rate of cluster destruction, as given by Eqs. (11) and (12).

Here, $s_i$ is cluster surface area given by:

$$s_i = 4\pi r_W^2 i^{2/3} = s_1 i^{2/3}, \tag{A3.2}$$

where $r_W$ is the Wigner-Zeitz radius of the material; $S$ is the saturation degree of the vapor; $n_i^e$ is equilibrium density of i-atomic clusters. When the spherical cluster approximation is accompanied with the Tolman correction for the surface tension, the equilibrium cluster density is given by:

$$n_i^e = n_1^e \exp\left(-\Theta \frac{i^{2/3}-1}{1+\alpha i^{-1/3}}\right). \tag{A3.3}$$

Here, $n_1^e$ is equilibrium vapor density; $\Theta$ is the non-dimensional surface energy defined in a traditional way (Warren and Seinfeld 1984):

$$\Theta = \frac{\sigma s_1}{kT}. \tag{A3.4}$$

$\sigma$ is the surface tension of the flat surface, $\alpha$ is non-dimensional Tolman length defined as:

$$\alpha = \frac{2\delta_T}{r_W}. \tag{A3.5}$$

Considering the denominator in the expression (A3.1) a continuous function of the cluster size $i$, and defining $f$ as a logarithm of it:

$$f(i) = ln(s_i n_i^e S^i). \quad (A3.6)$$

The Taylor expansion of the function $f$ near its maximum can be substituted in the definition of the nucleation rate (1). Truncating the resulting expression after the quadratic term yields the following approximation for the nucleation rate:

$$J = \frac{v_{th}}{4} n_1 \left(\frac{f''(i^*)}{2\pi}\right)^{1/2} exp(f(i^*)). \quad (A3.7)$$

Here, $i^*$ is the critical cluster size corresponding to the maximum of $f$, and $f''$ denotes the second derivative of $f$.

The critical cluster size $i^*$ can be determined from putting the function $f$ to zero. Substitution of the cluster surface area and equilibrium density ((A3.2) and (A3.3)) in the expression (6) and taking the derivative yields the following expression for $f'$:

$$f'(i) = lnS - \Theta \frac{2}{3} \frac{i^{-1/3}}{1+\alpha i^{-1/3}} - \Theta \frac{1}{3} \alpha i^{-4/3} \frac{i^{2/3}-1}{(1+\alpha i^{-1/3})^2} \quad (A3.8)$$

Following the derivation of the CNT (Girshick and Chiu 1990), we will use the following simplification, $\Theta i^{*2/3} \gg 1$, accompanied by $2i^* + \alpha i^{*2/3} \gg \alpha$. These two expressions yield the following simplified expression for $f'$ near $i^*$:

$$f'(i) = lnS - \frac{\Theta}{3} \frac{2i^{-1/3}+3\alpha i^{-2/3}}{(1+\alpha i^{-1/3})^2}. \quad (A3.9)$$

Equating $f'$ to zero yields a quadratic equation for $i^{*-1/3}$. We are only interested in the positive root:

$$i^{*-1/3} = \frac{lnS}{\Theta} \frac{3}{1-\gamma+\sqrt{1+\gamma}}, \quad (A3.10)$$

where

$$\gamma = 3\alpha \frac{lnS}{\Theta} = 6 \frac{\delta_T}{r_W} \frac{lnS}{\Theta} \quad (A3.11)$$

is a non-dimensional parameter.

Note that if the Tolman correction is not used, i.e., $\gamma = \alpha = 0$, then expression (A3.10) reduces to the well-known definition of the critical size from the CNT:

$$i^{*-1/3} = \frac{3}{2} \frac{lnS}{\Theta}.$$

The second derivative of near the critical cluster size reads:

$$H''(i) = \frac{2\Theta}{9} i^{-\frac{4}{3}} \frac{1+2\alpha i^{-\frac{1}{3}}}{\left(1+\alpha i^{-\frac{1}{3}}\right)^3}. \quad (A3.12)$$

Substituting (A3.10) into (A3.8) and (A3.12) yields the following expressions for $exp(f(i^*))$ and $f''(i^*)$:

$$exp(f(i^*)) = s_1 n_1^e \left(\frac{\Theta}{lnS} \frac{1-\gamma+\sqrt{1+\gamma}}{3}\right)^2 exp\left(\left(1 - \frac{\gamma}{1+\sqrt{1+\gamma}}\right)\Theta\left(1 - \frac{\Theta^2}{lnS^2} \frac{(1-\gamma+\sqrt{1+\gamma})^2(2-\sqrt{1+\gamma})}{27}\right)\right), \quad (A3.13)$$

$$f''(i^*) = 18 \frac{\ln S^4}{\Theta^3} \frac{1+\gamma+\sqrt{1+\gamma}}{\left(1+\sqrt{1+\gamma}\right)^3 \left(1-Y+\sqrt{1+\gamma}\right)^2}. \tag{A3.14}$$

Introducing another notation,

$$\lambda = \sqrt{1+\gamma} = \sqrt{1 + 6 \frac{\delta_T}{r_W} \frac{\ln S}{\Theta}}, \tag{A3.15}$$

allows to simplify these expressions:

$$\exp(f(i^*)) = s_1 \left(\frac{\Theta}{\ln S} \frac{(1+\lambda)(2-\lambda)}{3}\right)^2 n_1^e \exp\left((2-\lambda)\Theta \left(1 - \frac{\Theta^2}{\ln S^2} \frac{(1+\lambda)^2 (2-\lambda)^3}{27}\right)\right), \tag{A3.16}$$

$$f''(i^*) = 18 \frac{\ln S^4}{\Theta^3} \frac{\lambda}{(1+\lambda)^4 (2-\lambda)^2}. \tag{A3.17}$$

Substituting (A3.16) and (A3.17) into (A3.7) and taking into account that $n_1 = n_1^e S$ yields the following expression for the nucleation rate:

$$J = \frac{v_{th}}{4} s_1 n_1^{e2} S \left(\frac{\Theta}{\pi}\lambda\right)^{1/2} \frac{2-\lambda}{3} \exp\left((2-\lambda)\Theta \left(1 - \frac{\Theta^2}{\ln S^2} \frac{(1+\lambda)^2 (2-\lambda)^3}{27}\right)\right). \tag{A3.18}$$

Here, $\lambda$ is a non-dimensional parameter defined in (A3.15).

Noteworthy that in the case of constant surface tension (i.e., when the Tolman correction is not employed, $\delta_T = 0$, and $\lambda = 1$) expression (A3.18) reduces to a known formula for the nucleation rate from Girshick and Chiu (1990):

$$J = \frac{v_{th}}{12} s_1 n_1^{e2} S \left(\frac{\Theta}{\pi}\right)^{1/2} \exp\left(\Theta \left(1 - \frac{4}{27} \frac{\Theta^2}{\ln S^2}\right)\right) \tag{A3.19}$$

**Appendix 4. Benchmarking the cluster Gibbs free energy data**

The method of calculating the Gibbs free energies presented in Section 3.1 and the free energy values were benchmarked by using them to calculate an equilibrium cluster size distribution in a sub-saturated and close-to-saturated vapor and comparing it to the results of the cluster size distribution directly observed in the MD modeling for the same conditions, using the same classical inter-atomic potentials (Fig. A2).

MD simulations were run first, starting with a monomer vapor of specified density in a fixed box with periodic boundary conditions. Fixed temperature was maintained using a Langevin thermostat (corresponding to the NVT type of MD modeling) allowing for atoms to merge into clusters. The thermostat was efficiently withdrawing extra energy released from binding of atoms as they merge into clusters. Each of these simulations was continued until a steady-state cluster size distribution was obtained. The equilibrium density of individual vapor atoms left in the end of MD simulation was used as an input to analytically calculate equilibrium cluster densities based on the Gibbs energies by applying Eq. (13) sequentially to all cluster sizes, starting from dimers ($i=2$).

Three initial vapor densities were used in three separate MD simulations, $3.7 \times 10^{23} \ m^{-3}$, $2 \times 10^{24} \ m^{-3}$, and $5 \times 10^{24}$, resulting in these final vapor densities of $1.6 \times 10^{23} \ m^{-3}$, $4.1 \times 10^{23} \ m^{-3}$, and $5.8 \times 10^{23} \ m^{-3}$ respectively. These densities correspond to final vapor saturation degrees of approx. 0.4, 1, and 1.5 respectively, calculated using the equilibrium vapor density of $4 \times 10^{23} \ m^{-3}$, as identified in Section 3.2. Total number of atoms in the simulation box was $10^5$ in all three simulations, the size of the box was different (but fixed in each of the simulations) to obtain vapor density required.

Temperature of 2800 K was chosen for these tests, for two reasons. 1) Higher temperature is desirable since it implies exponentially higher cluster density in the saturated vapor needed for these tests, which means considerably shorter computation time. 2) At temperatures above 2800 K, the clusters start evaporating in the MD modeling in which the caloric curves were obtained, invalidating the caloric curves. In fact, clusters bigger than 10 atoms start evaporating at 2800 K and even lower temperatures, however, for the conditions considered in this section, they have low densities and can be neglected.

In each of the MD simulations, after a steady-state cluster size distribution was obtained, the Langevin thermostat was switched off (changing the simulation type from NVT to NVE). This was done to verify that the cluster size distribution is not distorted by the thermostat in any way. We have checked that, after the thermostat had been switched off, the temperature of the system did not change, neither did the cluster size distribution, verifying that a true equilibrium state of the system was obtained.

The comparison between the cluster size distributions obtained in the MD modeling and using the Gibbs free energies of cluster formation from the previous section are presented in Fig. A2. Only cluster sizes represented by more than 10 clusters (time-averaged) in the MD simulation boxed were plotted; bigger clusters with lower densities were omitted due to high statistical errors. Note that, the distribution in the highest density case with S > 1 is, strictly speaking, not steady-state. With S > 1, there is a finite cluster size at which the cluster Gibbs free energy peaks and then monotonically decreases for bigger clusters. These big clusters are energetically favorable and should form consuming monomers until the saturation degree S drops below 1. However, this critical size is quite big (tens of atoms) and the associated energy barrier is very high. Clusters can't overcome this barrier within MD-tractable timeframes. In other words, the big clusters are inaccessible in this simulation and can be considered non-existent. In this regard, the cluster distributions that we show in this case are indeed equilibrium distributions for clusters smaller than 10 atoms given that bigger clusters are not present.

This good agreement between the results confirms the accuracy of our method of determining Gibbs free energies of cluster formation and suggests that these energies can be used an input for a kinetic solver by Khrabry et al. (2024) of vapor condensation and cluster growth, where they can be used to calculate cluster evaporation rates. The kinetic solver equipped with these Gibbs energy data should be as accurate as classical MD for Fe clusters, and, apparently, much faster. We apply this solver to model Fe vapor condensation in Section 3.3.

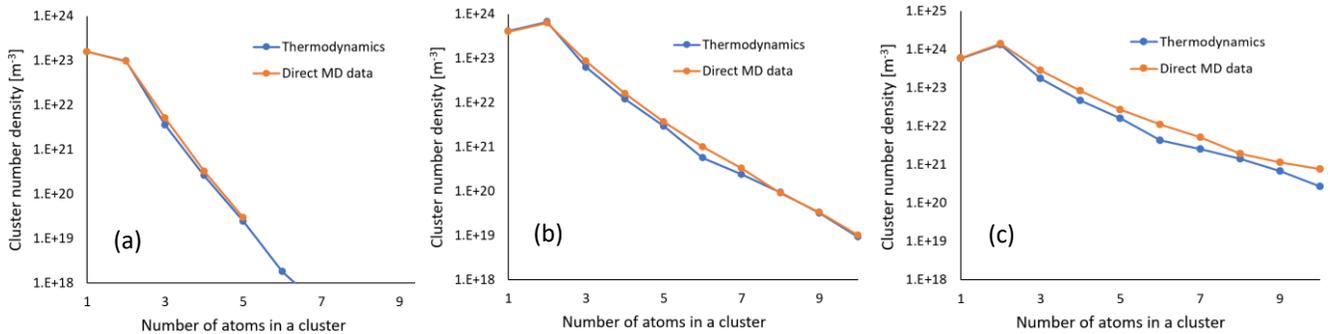

Fig. A2. Equilibrium cluster size distributions (starting from vapor atoms) at 2800 K for vapor densities of $1.6 \times 10^{23}\ m^{-3}$ (a), $4.1 \times 10^{23}\ m^{-3}$ (b), and $5.8 \times 10^{23}\ m^{-3}$ (c) obtained from thermodynamic considerations using the Gibbs free energy values presented in Section 3.1 and using direct MD modeling.

**Acknowledgments**

This work was funded by Princeton University.
**References**

Aktürk, A., and A. Sebetci (2016). BH-DFTB/DFT calculations for iron clusters, *AIP Advances*, 6, 055103; https://doi.org/10.1063/1.4948752

Bain, A. (2024). Recent advances in experimental techniques for investigating aerosol surface tension. *Aerosol Science and Technology*, 1–24. https://doi.org/10.1080/02786826.2024.2373907

Bakhtar, F., J. B. Young, A. J. White, and D. A. Simpson (2005). Classical Nucleation Theory and Its Application to Condensing Steam Flow Calculations, *J. Mech. Eng. Sci.*, 219, 1315, https://doi.org/10.1243/095440605X8379

Bergeron, D. E., A. W. Castleman, T. Morisato, and S. N. Khanna (2004). Formation of Al 13 I -: Evidence for the Superhalogen Character of Al 13. Science, 304(5667), 84–87; https://doi.org/10.1126/science.1093902

Bilodeau, J. F. and P. Proulx (1996). A Mathematical Model for Ultrafine Iron Powder Growth in a Thermal Plasma. *Aerosol Science and Technology*, 24 175–189; https://doi.org/10.1080/02786829608965362

Blokhuis, E. M., and J. Kuipers (2006). Thermodynamic expressions for the Tolman length, *J. Chem. Phys.*, 124, 074701; https://doi.org/10.1063/1.2167642

Bussi, G., D. Donadio, and M. Parrinello (2007). Canonical sampling through velocity rescaling, *J. Chem. Phys.*, 126, 014101; https://doi.org/10.1063/1.2408420

Castleman, A. W. and S. N. Khanna (2009). Clusters, Superatoms, and Building Blocks of New Materials. The Journal of Physical Chemistry C, 113(7), 2664–2675; https://doi.org/10.1021/jp806850h

Colombo, V., E. Ghedini, M. Gherardi, P. Sanibondi, and M. Shigeta (2012). A two-dimensional nodal model with turbulent effects for the synthesis of Si nano-particles by inductively coupled thermal plasmas, *Plasma Sources Sci. Technol.*, 21, 025001; http://doi.org/10.1088/0963-0252/21/2/025001

De Heer, W. A., W. D. Knight, M. Y. Chou, and M. L. Cohen (1987). Electronic Shell Structure and Metal Clusters. In Solid State Physics (Vol. 40, pp. 93–181). Elsevier; https://doi.org/10.1016/S0081-1947(08)60691-8

Ernzerhof, M., and J. P. Perdew (1998). Generalized gradient approximation to the angle- and system-averaged exchange hole. The Journal of Chemical Physics, 109, 3313–3320; https://doi.org/10.1063/1.476928

Finnis, M. W. and J.E. Sinclair (1984). A simple empirical N-body potential for transition metals, *Philosophical Magazine A*, 50, 45; https://doi.org/10.1080/01418618408244210

Frenkel, J (1955). *Kinetic Theory of Liquids*. New York: Dover Publications.

Frenklach, M. and S. J. Harris (1987). Aerosol dynamics modeling using the method of moments, *Journal of colloid and interface science*, 118, 252-261; https://doi.org/10.1016/0021-9797(87)90454-1

Frenklach, M. (2002). Method of moments with interpolative closure, *Chemical Engineering Science*, 57, 2229-2239; https://doi.org/10.1016/S0009-2509(02)00113-6

Friedlander, S. K. (1983). Dynamics of aerosol formation by chemical reaction, *Ann. N. Y. Acad. Sci.*, 404, 354; https://doi.org/10.1111/j.1749-6632.1983.tb19497.x
27